# Coherent Raman spectro-imaging with laser frequency combs

Takuro Ideguchi[1*], Simon Holzner[1*], Birgitta Bernhardt[1,3], Guy Guelachvili[2], Nathalie Picqué[1,2,3 †], Theodor W. Hänsch[1,3]

1. Max Planck Institut für Quantenoptik, Hans-Kopfermann-Str. 1, 85748 Garching, Germany
2. Institut des Sciences Moléculaires d'Orsay, CNRS, Bâtiment 350, Université Paris-Sud, 91405 Orsay, France
3. Ludwig-Maximilians-Universität München, Fakultät für Physik, Schellingstrasse 4/III, 80799 München, Germany

* These authors contributed equally to this work
† Corresponding author: nathalie.picque@mpq.mpg.de

**Optical spectroscopy and imaging of microscopic samples have opened up a wide range of applications [1-3] throughout the physical, chemical, and biological sciences. High chemical specificity may be achieved by directly interrogating the fundamental or low-lying vibrational energy levels of the compound molecules. Amongst the available prevailing label-free techniques, coherent Raman scattering has the distinguishing features of high spatial resolution down to 200 nm and three-dimensional sectioning. However, combining fast imaging speed and identification of multiple – and possibly unexpected- compounds remains challenging: existing high spectral resolution schemes require long measurement times to achieve broad spectral spans. Here we overcome this difficulty and introduce a novel concept of coherent anti-Stokes Raman scattering (CARS) spectro-imaging with two laser frequency combs. We illustrate the power of our technique with high resolution (4 $cm^{-1}$) Raman spectra spanning more than 1,200 $cm^{-1}$ recorded within less than 15 µs. Furthermore, hyperspectral images combining high spectral (10 $cm^{-1}$) and spatial (2 µm) resolutions are acquired at a rate of 50 pixels/s. Real-time multiplex accessing of hyperspectral images may dramatically expand the range of applications of nonlinear microscopy.**

With the advent of ultra-short pulse lasers and advanced photonics tools, coherent anti-Stokes and stimulated Raman scattering have evolved [4] into versatile tools for non-destructive chemically-selective diagnostics in complex systems, as encountered e.g. in material science, nanoscience, combustion phenomena or biology. A Raman band in a biological tissue may now be imaged at video rates [5,6]: with femtosecond lasers and fast lock-in detection techniques, the pixel dwell time may be as short as 100 ns. Spectral tailoring by means of pulse-shaping techniques enables three-color images to be acquired at such rates [7]. Alternatively, spectrally-chirping the femtosecond pulses enables fast tuning across the spectrum, useful in high resolution spectroscopy [8] or to optimize the contrast of single-color images [9]. Recently, frame-by-frame wavelength tuning across 300 $cm^{-1}$ has been reported at a rate of 30 frames/s [10].

Here we take a significantly different approach to Raman spectro-imaging and we present a simple concept of coherent anti-Stokes Raman scattering spectroscopy, based on two laser frequency combs, for ultra-rapid and highly multiplex spectral measurements. Initially conceived for frequency metrology [11], laser frequency combs are now becoming enabling tools for the rapid and sensitive acquisition of molecular linear absorption spectra [12-16] over a vast spectral span. Since the intensity of ultra-short pulses can easily be large enough to produce some nonlinear transient response in the observed medium, frequency combs may be harnessed for broad spectral bandwidth nonlinear molecular spectroscopy and imaging. This potential has not been exploited yet, except in a recent proof-of-principle demonstration [17] of stimulated Raman scattering dual-comb spectroscopy involving a complex set-up with three different laser systems.

CARS is a nonlinear four-wave mixing process, which is coherently driven when the energy difference of a pump and a Stokes laser beams is resonant with a Raman active molecular transition. Scattering off the probe beam provides the read-out through the generation of a high-frequency-shifted anti-Stokes signal enhanced by many orders of magnitude with respect to spontaneous Raman scattering. In our technique of dual-comb CARS spectroscopy, two femtosecond lasers with repetition frequencies $f+\delta f$ and $f$ irradiate a sample. In the time domain (Fig. 1a), a pulse from the first laser coherently excites a molecular vibration of period $1/f_{vib}$ longer than the pulse duration. The refractive index of the sample (Fig.1b) is thus modulated at the vibrational frequency. A pulse of the second laser probes the sample with a variable time separation $\Delta t$, that increases linearly from pulse pair to pulse pair. For simplicity, we illustrate a pulse that is short compared to the molecular vibration period [18]. If this second pulse arrives after a full molecular period $1/f_{vib}$, the vibration amplitude is increased and the back-action on the probe pulse is a spectral shift towards lower frequencies. If it arrives after half a period, the vibration amplitude is damped and the pulse experiences a shift towards higher frequencies. As long as the pulse separation $\Delta t$ remains shorter than the coherence time of the molecular oscillation, an intensity modulation of temporal frequency $f_{vib}\,\delta f/f$ is thus observed in the emitted anti-Stokes radiation. The two femtosecond lasers play a symmetric role: the sign of time separation $\Delta t$ between the pulses changes every $1/(2\delta f)$. In the frequency domain (Fig.1c,d), the two frequency comb generators produce an optical spectrum, which consists of several hundred thousands of perfectly evenly spaced spectral lines. Their frequencies may be described by:

$$f^{(1)}_m = m\,(f+\delta f) + f_{ceo} \quad\quad (1)$$
$$f^{(2)}_{m'} = m'\,f + f'_{ceo} \quad\quad (2)$$

where $m$, $m'$ are integers and $f_{ceo}$, $f'_{ceo}$ the carrier-envelope offset frequencies.

The frequency differences within each comb form regular combs themselves with vanishing carrier-envelope offset frequencies and line spacings of $f + \delta f$ and $f$, respectively. For instance, for comb 1 all pairs of lines with $m-n = k$ contribute to the same difference frequency $k\,(f+\delta f)$. Each of the difference frequency combs resonantly excites a molecular level of frequency $f_{vib}$ via Raman-like two-photon excitation, when a difference frequency comes close to $f_{vib}$, ie when $k \sim f_{vib}/f$. The excitations by the two combs interfere and modulate the molecular vibration at a beat note frequency $k\,\delta f = f_{vib}\,\delta f/f$. The two-photon excitation leads to a resonant enhancement of the third order nonlinear susceptibility observed via the anti-Stokes radiation. The intensity of the generated broadband anti-Stokes radiation is modulated at the beat note frequency $f_{vib}\,\delta f/f$. When several vibrational levels ($f_{vib1}$, $f_{vib2}$,…) are excited, the composite modulation contains all the beating frequencies ($f_{vib1}\,\delta f/f$, $f_{vib2}\,\delta f/f$…) representative of the involved levels. The Raman excitation spectrum is revealed by Fourier transformation of the intensity recorded versus time. The spectrum is mapped in the radio-frequency domain by the down-conversion factor $\delta f/f$ (typically of the order of $10^{-6}$). This grants ultra-rapid measurement time and efficient signal processing. Absolute calibration of the Raman-shifts is achieved by dividing the radio-frequencies by the down-conversion factor, which is easy to measure accurately. The carrier-envelope offsets cancel and do not have to be measured or controlled. This notably simplifies the experimental implementation and the calibration procedure. Similar modulation transfer phenomena had already been exploited in experiments using a single femtosecond laser and a phase-modulation pulse-shaper [19] or a Michelson interferometer [20-22]. However, the measurement times were fundamentally limited either by the sweep period of the phase modulation or by the mechanical motion in the Michelson interferometer. Our motionless frequency-comb-based technique results in dramatically shortened - more than a thousand fold- acquisition times, as well as in spectral resolution and spectral span only limited by the measurement time and the spectral bandwidth of the fs lasers.

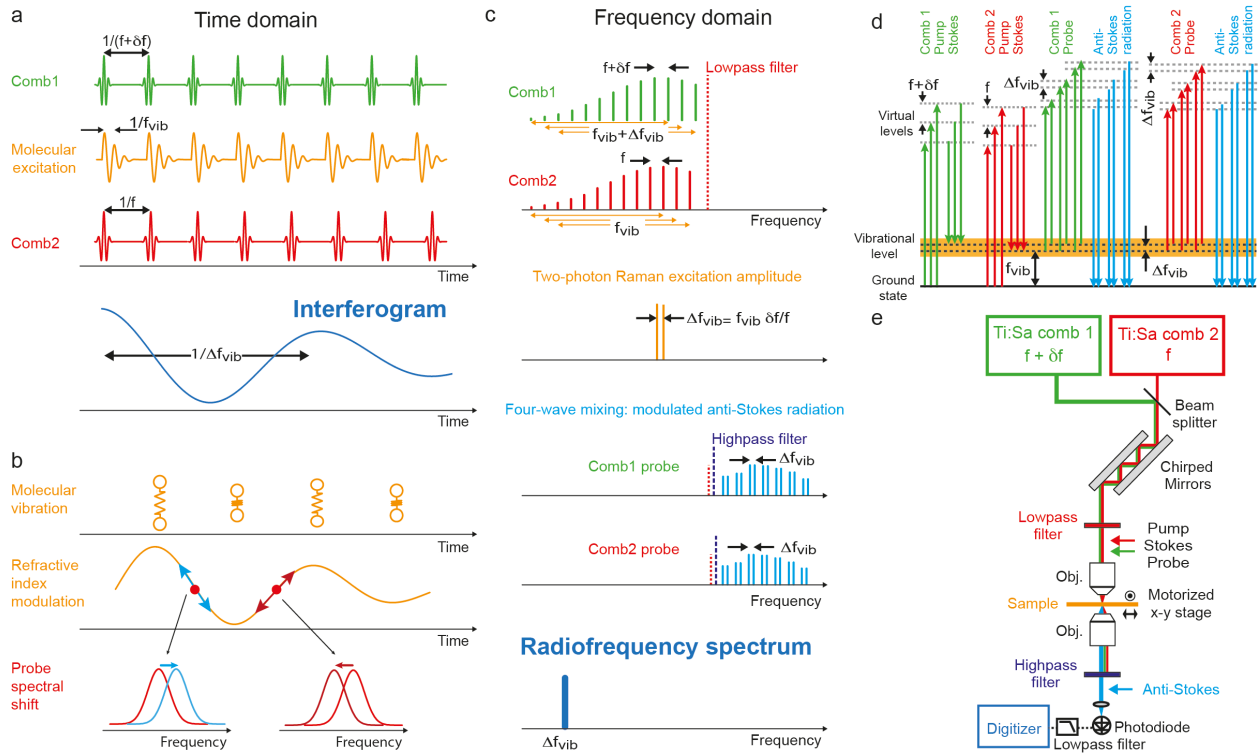

**Figure 1. Principle of dual-comb CARS spectroscopy**
Only one vibrational level is assumed to be within the spectral bandwidth of the laser combs. $\Delta f_{vib}$ stands for $f_{vib}\, \delta f/f$.
(a) Time domain representation of dual-comb CARS spectroscopy. The train of pulses of laser frequency comb 1 periodically excites the molecular vibration, which is probed by the pulses of laser frequency comb 2 with linearly increasing time delay. The filtered intensity modulation of the resulting anti-Stokes radiation provides the interferogram. The two combs play a symmetric role. In the figure, only the situation where the delay between the pulses of comb 2 and those of comb 1 is positive is displayed.
(b) When the probe pulse is short compared to the molecular oscillation (impulsive stimulated Raman scattering), the refractive index modulation of the sample -induced by the pump/Stokes beam- shifts the probe spectrum alternatively towards lower and higher frequencies.
(c) Frequency domain representation of dual-comb CARS spectroscopy. The two frequency combs modulate the excitation amplitude of the molecular vibration with a frequency $\Delta f_{vib}$. Such modulation is then transferred by the combs to the anti-Stokes radiation. For simplicity, the Raman excitations are represented as narrow lines.
(d) Energy level diagram, illustrating the four-wave mixing that leads to intensity-modulated anti-Stokes radiation.
(e) Experimental set-up, as described in the methods summary.

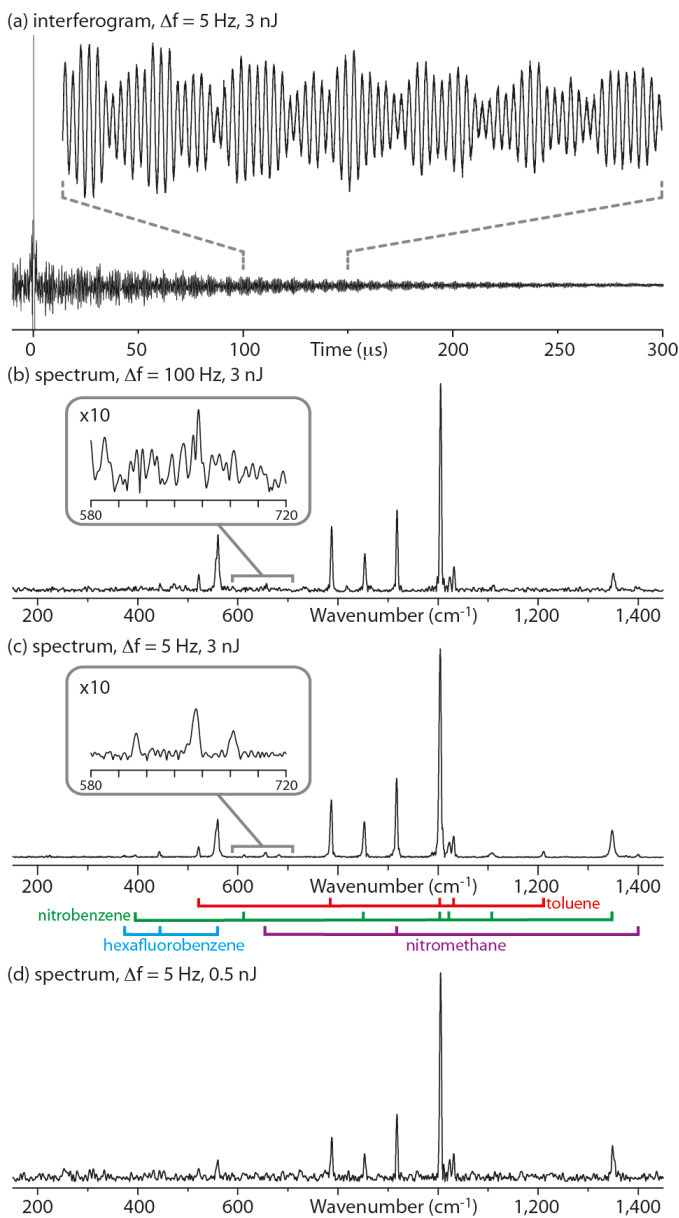

**Figure 2. Dual-comb CARS high-resolution spectroscopy of a mixture of liquid chemicals.**
(a) Interferogram showing the non-resonant interference signal around the zero time-delay and the interferometric modulation of the vibrational transitions of (c) (also in the zoomed part).
(b) Dual-comb CARS spectrum measured within 14.8 μs at 4 cm$^{-1}$ apodized resolution and with an energy per pulse of 3 nJ.
(c) Dual-comb CARS spectrum measured within 295.5 μs at 4 cm$^{-1}$ apodized resolution and with an energy per pulse of 3 nJ.
(d) Dual-comb CARS spectrum measured within 295.5 μs at 4 cm$^{-1}$ apodized resolution and with an energy per pulse of 0.5 nJ.
The insets graphs in (b) and (c) magnify ten fold the vertical scale. The number of individual spectral elements (defined as spectral span divided by resolution) in (b),(c) and (d) is 300.

Figure 1e displays the experimental set-up, detailed in the methods summary. It is similar to that used in dual-comb absorption spectroscopy [12-14] except for the dispersion management and spectral filtering to isolate the CARS signal from the comb beams. The time-domain interference signal -the interferogram- is periodic. Every $1/\delta f$, a strong burst mostly contains the non-resonant four-wave mixing signal resulting from the interference between the overlapping pulses of the two combs. A reproducible modulation (Fig. 2a), due to the CARS signal only, follows the burst and has a duration proportional to the coherence time of the sample transitions. A time-windowed portion of the interferogram, which excludes the interferometric non-resonant contribution, is Fourier transformed. The width of the window is determined according to the desired spectral resolution. The resulting spectra (Fig. 2b,c,d) span Raman-shifts from 200 cm$^{-1}$ up to 1,400 cm$^{-1}$. Importantly, the non-resonant background, which strongly lowers the sensitivity of CARS, is entirely suppressed, as in other specific CARS schemes [19-22].

Our ultra-rapid acquisition times are illustrated with three spectra at an apodized resolution of 4 cm$^{-1}$, recorded with δf =100 Hz (Fig. 2b) or 5 Hz (Fig. 2c,d). The sample is a mixture of hexafluorobenzene, nitrobenzene, nitromethane and toluene in a 5-mm-long cuvette. The measurement times are 14.8 μs (Fig. 2b) and 295.5 μs (Fig. 2c,d). Under constant clock rate sampling, the signal-to-noise ratio improves linearly with the inverse of the difference in repetition frequencies, culminating at 1,000 for the most intense blended line of toluene and nitrobenzene in Fig. 2c.

Moreover, we demonstrate spectro-imaging using a capillary plate (25-μm diameter holes, thickness: 500 μm) filled with a mixture of hexafluorobenzene, nitromethane and toluene as sample. A 45x45 μm$^2$ hyperspectral image is acquired: an interferogram is measured within 12 μs for each pixel and leads to a spectrum at an apodized resolution of 10 cm$^{-1}$. The total time of the experiment, 40.5 s, is limited by the refresh rate of the interferograms, while the entire sampling time of the interferograms is only 24.3 ms. After Fourier transformation, a spectral hypercube (Fig. 3) is obtained.

Our experimental concept demonstrates an intriguing potential for the rapid acquisition of broad spectral bandwidth high-resolution spectra and hyperspectral images of vibrational transitions. It can be further improved. A more sophisticated management of the dispersion, particularly accounting for third order dispersion, would result in an enhancement of the signal to noise ratio. Spectral broadening with nonlinear fibers or few-cycle oscillators will expand the spectral span. The implementation of fast lock-in detection schemes [6] might further decrease the noise level. The refresh rate of the interferograms could be increased a thousand fold with combs with a line spacing that matches the desired resolution. Chip-scale micro-resonators [23] or electro-optic modulators [24] may offer a route towards such achievement. For microscopy experiments, a straightforward way to speed-up the mapping process is to scan the laser beam with a galvanometric mirror rather than the sample. More interestingly, high-speed cameras (>1 Mfps) offer intriguing perspectives for real-time hyperspectral dual-comb CARS imaging.

Over the past years, several powerful schemes of CARS spectroscopy and imaging have been successfully exploited. They are foreseen to significantly expand the versatility of our technique along several directions. To give a few examples, surface-enhanced [25,26] dual-comb CARS may result in dramatically improved sensitivities. Raman optical activity [27] studied by dual-comb CARS might provide new insights on the structure and kinetics of chiral molecules, e.g. during asymmetric chemical reactions. Sub-wavelength spatial resolution in our imaging experiments might be achieved either in the near-field e.g. at a metal tip [28] or in the far-field e.g. by state depletion [29].

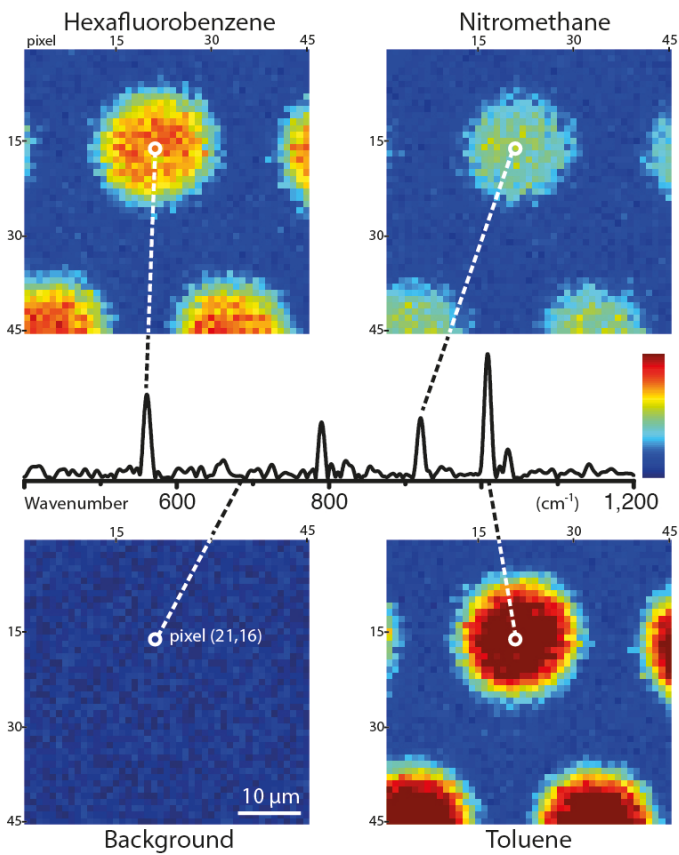

**Figure 3. Hyperspectral image of a capillary plate with holes filed with a mixture of hexafluorobenzene, nitromethane and toluene.**
Each of the 2025 pixels of the hyperspectral cube corresponds to a spectrum at 10 cm$^{-1}$ apodized resolution measured within 12μs at fixed spatial location and provides the spectral signature of compounds present in this part of the sample. The spectrum shown in the center of the figure corresponds to pixel (21,16). Each spectral element of the hyperspectral cube may be plotted as an image similar to the four ones displayed in the figure and provides spatial quantitative identification of a given compound. An image corresponds to intensity for all pixels at a fixed wavenumber. This carries information about the spatial distribution of a compound with a distinguishable spectral signature at that wavenumber.

**Methods summary:**

Two Titanium-sapphire lasers (Femtolasers, Synergy20 UHP) emit 20 fs pulses centered at 12,580 cm$^{-1}$ with energies up to 13 nJ. Both have adjustable repetition frequencies of about 100 MHz. The beams of the two lasers are combined on a beam-splitter and a chirped mirror compressor (Layertec) compensates for the second-order dispersion induced by the optical components of the set-up. Spectral filtering is applied to improve the signal to background ratio. A low-frequency pass filter (Chroma Technology, ET750LP, cutoff: 13,330 cm$^{-1}$) before the sample and a high-frequency pass filter (Omega Optical Inc., 3RD740SP, cutoff: 13,510 cm$^{-1}$) after the sample isolate the CARS signal that is generated by the sample after proper focusing with a lens or a microscope objective. The spectral span is limited thus on the low-energy side by the optical filters and on the high-energy side by the spectral bandwidth of the fs lasers. The anti-Stokes radiation is forward-collected and focused onto a silicon photodiode with a frequency bandwidth of the order of 100 MHz. The electric signal is low-pass filtered to 50 MHz to avoid aliasing. It is then amplified and digitized with a 180 MSamples/s, 16-bit data acquisition board (Alazartech, ATS9462).

The data of Fig. 2 have been recorded with orthogonal linear polarizations of the two laser beams. This decreases the interferometric non-resonant background, while the fast depolarization of the sample maintains the anti-Stokes signal strong. For the spectra of Fig. 2b,c, the focusing optics consists in 20-mm focal length lenses and requires an amount of dispersion compensation of -600 fs$^2$. The pulse energy at the sample is 3 nJ. The spectrum of Fig. 2d is measured with a 8-mm focal length focusing lens, a pulse energy of 0.5 nJ and an avalanche photodetector. To record the hyperspectral images (Fig. 3), the difference in repetition frequencies of the two combs is set to 50 Hz and a microscope objective (Olympus LCPLN20XIR) focuses the beams onto the sample, with a beam diameter of 1.9 µm and a Rayleigh length of 3.4 µm. The pulse energy at the sample is 3.8 nJ and a second-order dispersion of -3,000 fs$^2$ is compensated for. The sample is mounted on a motorized x-y platform that raster scans across the sample at 1 µm steps.

**Acknowledgements**
We warmly thank Peter Hommelhoff, Martin Schulze and Wolfgang Schweinberger for the loan of optical components and Arthur Hipke for experimental support.
Research conducted in the scope of the European Laboratory for Frequency Comb Spectroscopy. Support by the Max Planck Foundation, the Munich Center for Advanced Photonics, Eurostars and the European Research Council (Advanced Investigator Grant 267854) are acknowledged.